\documentclass[twocolumn,aps,superscriptaddress,preprintnumbers]{revtex4-1}

\usepackage{xcolor}
\usepackage{graphicx}
\usepackage{dcolumn}
\usepackage{bm}
\usepackage{hyperref}

\usepackage{float,amsmath,amssymb,amsfonts,latexsym,color}

\usepackage{hyperref}

\hypersetup{colorlinks=true,       
            linkcolor=blue,        
            citecolor=blue,        
            filecolor=blue,       
            urlcolor=blue,         
            runcolor=blue
}

\begin{document}

\title{Detecting zero-point fluctuations with stochastic Brownian oscillators}

\author{Adrian E. Rubio Lopez}
\email{adrianrubiolopez0102@gmail.com}

\author{Felipe Herrera}

\affiliation{Department of Physics, Universidad de Santiago de Chile, Av. Victor Jara 3493, Santiago, Chile.}
\affiliation{Millennium Institute for Research in Optics, Concepci\'on, Chile.}

\date{\today}           

\begin{abstract}
High-quality quantum oscillators are preferred for precision sensing of external physical parameter because if the noise level due to interactions with the environment is too high, metrological information can be lost due to quantum decoherence. On the other hand, stronger interactions with a thermal environment could be seen a resource for new types of metrological schemes. We present a general amplification strategy that enables the detection zero-point fluctuations using low-quality quantum oscillators at finite temperature. We show that by injecting a controllable level of multiplicative frequency noise in a Brownian oscillator, quantum deviations from the virial theorem can be amplified by a parameter proportional to the strength of the frequency noise at constant temperature. As an application, we suggest a scheme in which the virial ratio is used as a witness of the quantum fluctuations of an unknown thermal bath, either by measuring the oscillator energy or the heat current flowing into an ancilla bath. Our work expands the metrological capacity of low-quality oscillators and can enable new measurements of the quantum properties of thermal environments by sensing their zero-point contributions to system variables.
\end{abstract}
\maketitle

Brownian motion of classical and quantum oscillators subject to dissipation and noise fluctuations is a paradigmatic model with direct fundamental \cite{LevinPRE,HeatTransferQBM,BrownianReview,GravityClassicalDiffusion}  and experimental \cite{TracyHighQ} implications in several areas of physics, from states tomography and spectroscopy \cite{BlattReview,AspelmeyerReview} to tests on gravitation \cite{ReviewGrav}. Dissipation and noise result from the interaction of the system oscillator with its environment, often assumed as a large set of degrees of freedom that are not controlled but influence the dynamics of the system \cite{Agarwal1971}. Strong interactions with a thermal environment are avoided in metrological schemes, as they reduce the sensitivity of system properties to external parameters such as external fields. This is critical in quantum sensing, as the metrological information content is severely limited by environment-induced decoherence \cite{Escher2011}. 

Implementing high quality (high-$Q$) oscillators that are well-isolated from the environment is essential for metrological schemes that map unknown electric and magnetic fields to variations in the oscillator frequency \cite{McCormick2019}. Frequency shifts are more difficult to measure with noisier low-$Q$ oscillators. On the other hand, stronger dependence of the oscillator observables on the detailed structure of the environment can open new opportunities for obtaining information from the surroundings. For a system interacting with environment formed by a mediator field coupled to a source system, studying the system’s effective dynamics could reveal fundamental aspects of the mediator field even without having access to its full description. This might be of great interest for deciding, for instance, whether gravity is quantum or not. Quantum mechanical proposals to address this problem include full spectroscopy of the environment \cite{Paz-Silva2017},  generation of entanglement through a mediator field \cite{ReviewGrav,SpinWitness} or studying decoherence due to environmental fields \cite{Marletto1,Gundhi}. However, witnessing the nature of an environment might be accessible without appealing to intrinsic quantum features (entanglement, decoherence), as it occurs with Casimir forces \cite{ReviewWoods,ReviewBuhmann}, where zero-point fluctuations manifest macroscopically \cite{MilonniQV}.

For classical harmonic oscillators that are completely isolated from their environments, the virial theorem establishes that the average kinetic energy $\langle K\rangle$ equals the average potential energy $\langle V\rangle $ \cite{Goldstein}. The virial ratio $\mathcal{R}\equiv \langle K\rangle/\langle V\rangle$ is also equals to one for quantum harmonic oscillators that undergo unitary dynamics \cite{Hirschfelder1960}. Environment-induced deviations of $\mathcal{R}$ from unity have only recently been explored in the quantum regime from a formal perspective \cite{Bialas2018,VirialQuantumBrownian}, but a connection between the virial ratio and experimentally accessible oscillator variables has not been developed yet.

\begin{figure}[t]
\includegraphics[width=0.4\textwidth]{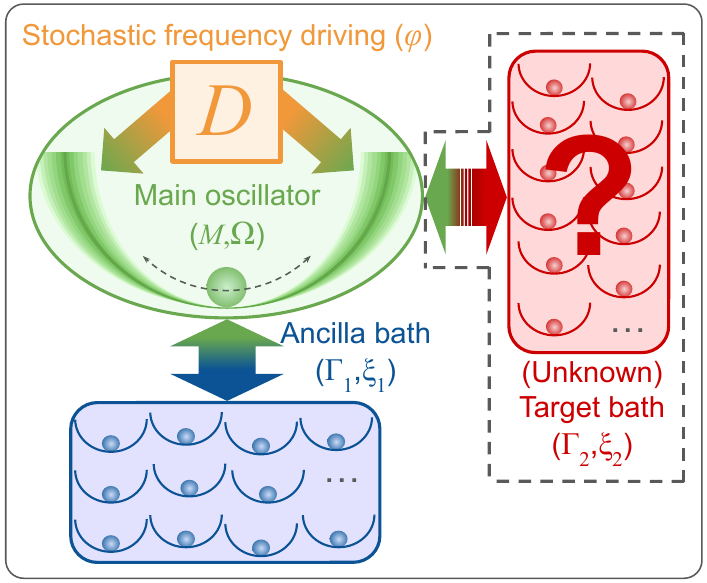}
\caption{Scheme of the scenario. A nonequilibrium Brownian oscillator of mass $M$ and frequency $\Omega$ is subjected to white frequency noise (given by the stochastic variable $\varphi$ and characterized by the strength $D$). The oscillator is coupled to two Ohmic baths characterized by damping constants $\Gamma_{1}$ and $\Gamma_{2}$, stochastic force $\xi_{1,2}$ and large cutoff frequency ($\Omega,\Gamma_{1,2}\ll\Omega_{\rm C}$). Ancilla bath is assumed to introduce strong dissipation ($\Gamma_{1}\lesssim\Omega$). The nature of the target bath is assumed unknown.
}
\label{FigScheme}
\end{figure}

In this Letter we show that a low-$Q$ Brownian oscillator that is subject to a frequency noise at low temperatures can be used to probe environmentally-induced changes in the virial ratio $\mathcal{R}$, either by measuring the oscillator energy $E$ or heat currents $J$ between system and the environment. Deviations from $\mathcal{R}=1$ are shown to scale with the tunable magnification factor $\mathcal{W}=QD\Omega/(1-QD\Omega)$, where $D$ is the frequency noise strength and $\Omega$ the oscillator natural frequency. Such magnification can be exploited for precision measurements. At low temperatures, the proposed scheme is shown to represent a feasible approach for measuring zero-point fluctuations. We also propose a two-bath protocol for probing the quantum nature of fluctuations of an unknown thermal bath that couples to the frequency-driven Brownian oscillator,  by measuring the oscillator energy or the heat current flow between the system  and an ancilla bath, as illustrated in Fig. \ref{FigScheme}.

{\it Stochastic Brownian oscillator:} We consider a harmonic oscillator whose frequency is randomly driven and simultaneously interacts with one or more thermal environments. Frequency noise is given by $\Omega(t)=\Omega[1+\varphi(t)]^{1/2}$, with white noise amplitude $\varphi(t)$ of zero mean and second moment $\langle\varphi(t)\varphi(t')\rangle=2D\delta(t-t')$. $D$ is the noise strength (in Hz$^{-1}$). Higher orders cumulants are $\langle\langle\varphi(t_{1})...\varphi(t_{n})\rangle\rangle=2^{n}D_{n}\delta(t_{1}-t_{2})...\delta(t_{n-1}-t_{n})$ with $D_{n}$ taken as in Ref.~\cite{West1980}. Physically, frequency fluctuations could be implemented using laser intensity noise in levitated nanoparticles by optical tweezers \cite{LevitatedNanoparticleNovotny}, voltage noise in an ion traps \cite{BlattReview}, or spectral diffusion in single-molecule substrates \cite{Sarkar2024} (See Appendix \ref{app:MultiplicativeNoiseImplementation} for suggested physical implementations). We neglect any non-linear back-reaction associated with the injection of noise energy in the oscillator~\cite{LindenbergNonlinear}. For the analysis below, the oscillator can be classical or quantum mechanical. 

Thermal environments correspond to sets of harmonic oscillators in thermal states with temperature $T_{k}$, where $k$ labels different baths ($k=\{1,2\}$ in Fig. \ref{FigScheme}). Bath oscillators are linearly coupled to the main system oscillator (see Caldeira-Leggett model details in Appendix \ref{app:CaldeiraLeggett}). Eliminating the bath variables results in damping terms and stochastic fluctuation forces represented by the random variables $\xi_{k}$, acting on the main oscillator. The former are given by $(d/dt)[\int_{0}^{t}d\tau\tilde{\Gamma}_{k}(t-\tau)x(\tau)]$, {having $x(t)$ as the main oscillator amplitude and $\tilde{\Gamma}_{k}(t)$ the damping kernels which depend} on the bath cutoff frequency $\Omega_{\rm C}$ and the damping constants $\Gamma_{k}$.

The system dynamics is non-Markovian in general, but for large cutoff frequencies $\Omega_{\rm C}\gg\{\Omega,\Gamma_{1,2}\}$ we have $(d/dt)[\int_{0}^{t}d\tau\tilde{\Gamma}_{k}(t-\tau)x(\tau)]\approx 4\Gamma_{k}\dot{x}(t)$, which gives Markovian evolution. The resulting equation of motion for  $x(t)$ reads
\begin{equation}
\ddot{x}(t)+\Omega^{2}\left[1+\varphi(t)\right]x(t)+4\Gamma\dot{x}(t)=\frac{\xi(t)}{M},
\label{EqMotion}
\end{equation}
where $\Gamma=\Gamma_{1}+\Gamma_{2}$ and $\xi=\xi_{1}+\xi_{2}$, with stochastic force correlations given by $\langle\xi(t)\xi(t')\rangle=(1/2)[N_{1}(t-t')+N_{2}(t-t')]$ due to the independence of the baths (i.e., $\langle\xi_{1}(t)\xi_{2}(t')\rangle=0$). The  noise kernels (symmetrized correlations) are $N_{k}(t-t')=\langle\xi_{k}(t)\xi_{k}(t')+\xi_{k}(t')\xi_{k}(t)\rangle$, where the expectation value is over the bath ensemble. For quantum thermal baths, noise kernels are given by ($\hbar=1$)\cite{BreuerPett,CalzettaHu}:
\begin{equation}
N_{k}(t-t')=2\int_{0}^{+\infty}d\omega\mathcal{J}_{k}(\omega)\coth\left(\frac{\omega}{2k_{\rm B}T_{k}}\right)\cos\left[\omega(t-t')\right],
\label{NoiseKernel}
\end{equation}
where $\mathcal{J}_{k}(\omega)=(2M\Gamma_{k}\omega/\pi)f(\omega/\Omega_{\rm C})$ is the spectral density of the $k-$th bath, which we assume to be Ohmic with a Lorentzian cutoff function $f(x)=1/(1+x^{2})$. The thermal factor $\coth(\omega/[2k_{\rm B}T_{k}])=1+2\overline{n}_{k}(\omega)$ presents the sum of the contributions of the zero-point and the thermal fluctuations, with the latter associated to the Bose-Einstein distribution $\overline{n}$. The high-temperature limit for a given bath corresponds to $\Omega,\Gamma_{1,2}\ll\Omega_{\rm C}\ll k_{\rm B}T_{k}$, which allows the approximation 
$N_{k}(t-t')\approx 4M\Gamma_{k}k_{\rm B}T_{k}~\Omega_{\rm C}{\rm Exp}[-\Omega_{\rm C}(t-t')]$, i.e., exponentially-correlated noise. A sufficiently large cutoff frequency gives the classical $N_{k}(t-t')\rightarrow 4M\Gamma_{k}k_{\rm B}T_{k}\delta(t-t')$, corresponding to white noise. Comparing the latter form with the damping kernel in the large cutoff limit, the classical Fluctuation-Dissipation Relation (FDR) is verified. In this work, we consider the high-temperature limit as a definition of classical baths. This criteria relies on the fact that the classical limit neglects both the zero-point fluctuations and the (quantum) blackbody features of the occupation number $\overline{n}_{k}$.

{\it Steady-state energy distribution}: Equation (\ref{EqMotion}) without dissipation describes parametric resonances when the variable $\varphi(t)$ is replaced by harmonic driving ($\varphi(t)\rightarrow A_{D}\cos[\omega_{D}t]$), with some amplitude $A_D$ and driving frequency $\Omega_D$. For specific values of $\omega_{D}$, the oscillator undergoes exponential growth of $x(t)$ with time \cite{LandauLifshitz}. In our case, frequency driving is not harmonic but random, and white noise injects energy on the system at all frequencies. A sufficient amount of dissipation can compensate for this energy injection and lead the system to a stable steady state. Unbalanced energy injection can give unstable {dynamics}. Stability criteria for our case are discussed below. 

In the large cutoff regime, we follow the approach in Ref.\cite{West1980} for solving Eq.(\ref{EqMotion}) to obtain first and second order moments of $x$. Analytical expressions in the steady state can be derived under the assumption of white noise for $\varphi$, while preserving the quantum fluctuations on the bath, despite the main oscillator being formally treated as statistically classical (see Refs.\cite{GittermanPaper,GittermanBook} for approaches restricted to the classical case). From $\langle x^2\rangle $ and $\langle p^2\rangle$, the stationary energy $E$ of the frequency-driven oscillator can be written as
\begin{equation}\label{EnergyWithFreqNoise}
E=E_{0}\left[1+2\mathcal{W}/(1+\mathcal{R})\right],
\end{equation}
where $E_{0}=\langle K\rangle+\langle V\rangle$ is the energy of the undriven Brownian oscillator ($D=0$), $\mathcal{W}=QD\Omega/(1-QD\Omega)$ is the amplification factor, $Q=\Omega/4\Gamma$ is the quality factor of the main oscillator and $\mathcal{R}$ the virial ratio. The derivation of Eq. (\ref{EnergyWithFreqNoise}) is summarized in Appendix \ref{app:WestApproachForEnergy}.

Energy stability requires that $QD\Omega<1$. Within the space of stable solutions, the fact that the correction to the undriven energy $E_0$  scales with $D$, for fixed $Q$ and $\Omega$, opens the opportunity of probing the environment induced deviations of the classical virial ratio $\mathcal{R}=1$. Such deviations have been discussed \cite{VirialQuantumBrownian}, but feasible experimental schemes for detecting such deviations are not available. Equation (\ref{EnergyWithFreqNoise}) shows, for the first time, that by directly measuring the energy of a classical or quantum Brownian oscillator with tunable frequency noise strength $D$, bath-induced deviations from the virial theorem can be amplified in stationary measurements. Rewriting Eq. (\ref{EnergyWithFreqNoise}) as $E/E_0=1+\mathcal{F}\mathcal{W}$, with $\mathcal{F}=2/(1+\mathcal{R})$, suggests that by measuring $E$ with different driving strengths $D\sim \mathcal{W}$, deviations from the classical virial ratio could be found from the slope $\mathcal{F}\leq1$, with the equality holding in the classical limit $\mathcal{R}_{\rm cl}=1$. 

{\it Detecting deviations from the virial theorem:} To understand the regimes of $\mathcal{R}$ that could accessible with oscillator energy measurements, it is instructive to rewrite the stationary solution for $\langle K\rangle$ as (see Appendix \ref{app:WestApproachForEnergy} for the derivation)
\begin{equation}\label{KinAndPot}
\langle K\rangle = \frac{\Omega}{2}\int_{0}^{\infty}\frac{du}{\pi}\frac{u\coth(u/\widetilde{T})}{Q\left(\left[1-u^{2}\right]^{2}+[{u}/{Q}]^2\right)} u^{2}f\left(u/u_{\rm C}\right),
\end{equation}
where $\widetilde{T}=T/T_{0}$ is the ratio between the thermal energy ($k_{\rm B}T$) and the oscillator zero-point energy ($\Omega/2$), i.e.,  $T_{0}\equiv\Omega/(2k_{\rm B})$; and $u_{\rm C}\equiv \Omega_{\rm C}/\Omega$ is a cutoff parameter. The potential energy $\langle V\rangle$ has the same integral expression as $\langle K\rangle$ with the replacement $u^2 f(u/u_{\rm C})\rightarrow 1$ (see Appendix \ref{app:WestApproachForEnergy}). Only the kinetic energy needs to be regularized with a cutoff function mantaining the well-known dependence on $u_{\rm C}$ \cite{HanggiIngoldQBM}. Deviations from the classical {virial} theorem ($\mathcal{R}_{\rm cl}=1$) can be expected in the combined regime of low temperatures ($\widetilde{T}\lesssim 1$) and strong system-bath coupling (low $Q$). For high-temperatures [$\widetilde{T}\gg u_{\rm C}$ in Eq. (\ref{KinAndPot})] the classical ratio  $\langle K\rangle=\langle V\rangle\approx k_{\rm B}T/2$ holds for all system-bath coupling strengths, in agreement with the equipartition theorem. High-$Q$ oscillators satisfy {$\mathcal{R}=1$} with $\langle K\rangle=\langle V\rangle = (\Omega/4)\coth(1/\widetilde{T})$ at all temperatures. 

\begin{figure}[t]
\includegraphics[width=0.45\textwidth]{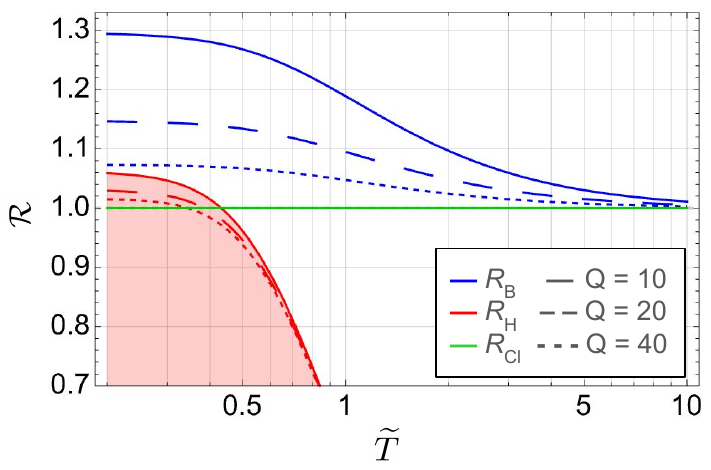}
\caption{Virial ratio $\mathcal{R}=\langle K\rangle/\langle V\rangle$ as a function of the normalized temperature $\widetilde{T}$ for an oscillator with quality factor $Q=10,20,40$ (solid, long dashed, short dashed) and a cutoff frequency for the bath $Q_{\rm C}=10^{3}$. The blue curves correspond to the full expression for a Brownian oscillator in the steady state under the influence of a quantum thermal bath, including zero-point fluctuations. Red curves correspond to the lower bound on $R$ obtained from applying the Heisenberg uncertainty principle (shaded area). The green solid line corresponds to the value of the ratio for a Brownian oscillator under the influence of a `classical' thermal bath.}
\label{FigVirial}
\end{figure}

Figure \ref{FigVirial} shows the virial ratio $\mathcal{R}$ for a Brownian oscillator ($\mathcal{R}_{\rm B}$) as a function of the reduced temperature $\widetilde T$, for different quality factors $Q$. The cutoff parameter is set to $u_{\rm C}=10^3$. As discussed above, the classical limit $\mathcal{R}_{\rm cl}=1$ is approached as the thermal energy exceeds the zero-point energy ($\widetilde{T}\gg 1$) and low-temperature deviations from the classical limit are suppressed as the coupling to the bath becomes weaker (higher $Q$ values). Up to $\sim 30\%$ deviations from {the classical limit are} expected at lower temperatures ($\widetilde{T}\sim 0.1$) for relatively lossy oscillators ($Q\sim 10$). We ruled out that the calculated deviations from {$\mathcal{R}_{\rm cl}=1$} are a cutoff-dependent artifact by comparing with the {fundamental} lower bound $\mathcal{R}_{\rm B}\geq\mathcal{R}_{\rm H}\,\equiv\,\Omega^{2}/16\langle V\rangle^{2}$, obtained from Heisenberg's uncertainty relation $\Delta p^{2}\Delta x^{2}\geq 1/4$, given that $\langle x\rangle=\langle p\rangle=0$. Since  $\mathcal{R}_{\rm H}$ only depends on the potential energy, it is effectively cutoff-independent in the large cutoff regime. While the Brownian oscillator model {gives deviations from the virial theorem that depend on the cutoff parameter, the Heisenberg bound guarantees that we still expect bath-induced deviations at very low temperatures. 

\begin{figure*}[t]
\includegraphics[width=\linewidth]{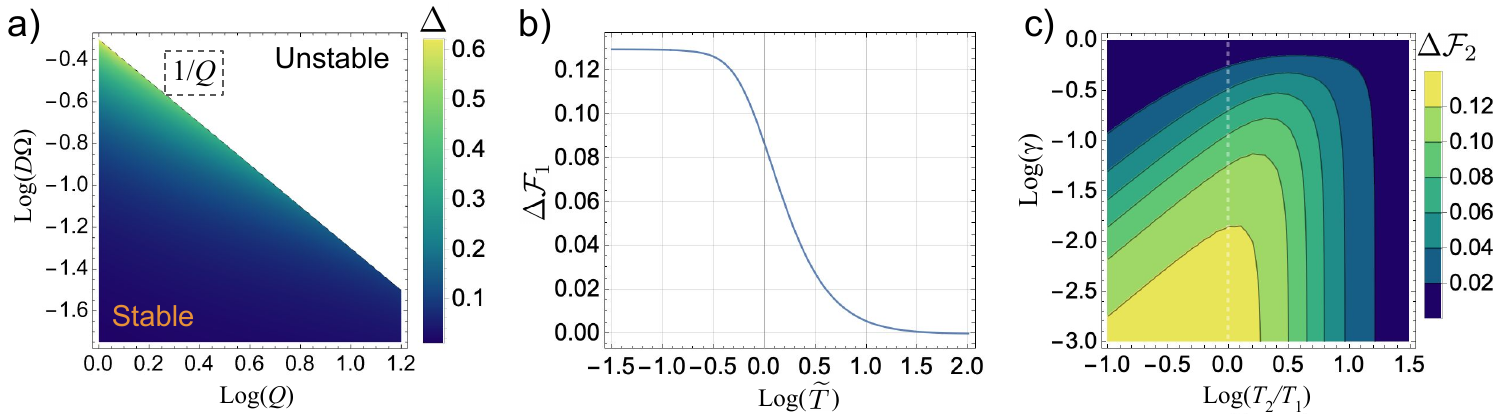}
\caption{(a) {Net energy deviation} $\Delta=\mathcal{W}(1-\mathcal{F})$ over the stability region of an oscillator at temperature $\widetilde{T}=1/4$ as a function of $D\Omega$ and $Q$, with $u_{\rm C}=10^{3}$. 
(b) Difference of the virial factor $\Delta\mathcal{F}_{1}=|\mathcal{F}_{\rm Q}-\mathcal{F}_{\rm cl}|$ as a function of the normalized temperature $\widetilde{T}$ for the single bath case when the bath is either quantum of classical, with $Q=10$ and $u_{\rm C}=10^{3}$.
(c) Variation of the virial factor $\Delta\mathcal{F}_{2}=|\mathcal{F}_{2,\rm Q}-\mathcal{F}_{2,\rm cl}|$ for the two baths scenario with temperatures $T_{1,2}$ and different damping rates such that the relative damping is defined $\gamma\equiv\Gamma_{1}/(\Gamma_{1}+\Gamma_{2})$. The variation is considered between {scenarios where the ancilla bath is fixed to be quantum and the target is either quantum or classical}. The dashed vertical line corresponds to thermal equilibrium ($T_{1}=T_{2}$), with $\widetilde{T}_{1}=1/4$, $Q=10$ and $u_{\rm C}=10^{3}$.}
\label{FigCombined}
\end{figure*}

As mentioned above, there is a limit to how strongly the oscillator can be driven without undergoing parametric amplification into unstable steady states. Figure \ref{FigCombined}(a) shows the magnitude of the {net energy deviation $\Delta=\mathcal{W}(1-\mathcal{F})$ when the bath is either classical or quantum}, expected for different $Q$-factors and dimensionless noise strength $D\Omega$, for stable configurations satisfying $QD\Omega<1$. Relatively large {net energy deviations $\Delta\sim 0.2$} are expected for $Q<10$ and $\varphi$-noise strength $\sqrt{2D\Omega}\sim 0.35$. This suggests that low-$Q$ oscillators are more suitable for observing larger variations of $\mathcal{F}$.
 
Deviations from classical virial theorem are larger at lower temperatures, because zero-point bath fluctuations become important. A manifestation of this is the strong dependence of $\mathcal{R}_{\rm B}$ with the bath's cutoff frequency $\Omega_{\rm C}$. This dependence is lost in the high temperature (classical) limit when the Bose-Einstein distribution cutoff $\overline{n}$ dominates over $\Omega_{\rm C}$. In other words,  by measuring the total oscillator energy $E$, {and accessing} $\mathcal{R}$, it is  possible to assess whether the thermal environment is quantum or classical. Figure~\ref{FigCombined}(b) shows the differences in slopes $\Delta\mathcal{F}_{1}=|\mathcal{F}_{\rm Q}-\mathcal{F}_{\rm cl}|$ between quantum and classical thermal baths, as a function of the dimensionless temperature $\widetilde T$. Quantum baths satisfy FDR via the quantum thermal factor $\coth(u/\widetilde{T})$, i.e., including both zero-point and blackbody contributions, and classical baths satisfy FDR through the high-temperature factor $\widetilde{T}/u$.  For both cases, we set $Q=10$ and $u_{\rm C}=10^{3}$. As expected, at higher temperatures zero-point contributions cannot be distinguished, but differences $\Delta \mathcal{F}$ of over 12$\%$ are predicted at low temperatures.

{\it Probing the {nature} of an unknown bath:} {Stochastic frequency} Brownian oscillators can be used as probes of a second bath (see Fig. \ref{FigScheme}). A known \emph{ancilla} bath {with tunable damping rate $\Gamma_1$ and temperature $T_1$} is coupled to the main oscillator, which in turn is coupled to a second \emph{target} bath, such that only the damping constant $\Gamma_2$ is previously known from the oscillator's relaxation dynamics without frequency noise. The temperature of the target bath ($T_{2}$) is only approximately estimated, so it is possible to look for differences in the virial ratio. The ancilla bath is quantum mechanical (i.e., {includes zero-point fluctuations and Bose-Einstein thermal fluctuations}), but the quantum or classical nature of the target bath is unknown. 

For non-equilibrium scenarios ($T_{1}\neq T_{2}$), a steady heat current is expected to flow between baths from higher to lower temperature, through the main oscillator \cite{HeatTransferQBM}. Frequency noise acts as an additional energy injection source, therefore contributing to the stationary heat current between baths. From the power balance expression $dE/dt=J_{1}+J_{2}+W$, where $W\equiv\Omega^{2}\langle\varphi xp\rangle$ is the energy injected by the frequency noise, we extend the methods in Refs. \cite{West1980,Kubo1963} to obtain the heat current $J_{k}$ from the driven oscillator to the $k$-th bath of the form (derivation in Appendix \ref{app:HeatCurrents})
\begin{equation} \label{HeatCurrentsFull}
J_{k}=J_{k}^{(0)}-4\Gamma_{k}E_{0}\mathcal{W}\mathcal{F},
\end{equation}
where the first term $J_{k}^{(0)}(T_1,T_2)$ is the standard heat current without frequency noise, and the second term is a noise-induced correction, which again scales linearly with the magnification factor $\mathcal{W}$ and the virial factor $\mathcal{F}$. The virial factor is a function of the properties of the two baths, i.e., $\mathcal{F}=\mathcal{F}(T_{1},T_{2})$. The same occurs for the energy $E_{0}=E_{0}(T_{1},T_{2})$. The quality factor $Q$ is defined over the total damping $\Gamma=\Gamma_{1}+\Gamma_{2}$.

Equation (\ref{HeatCurrentsFull}) suggests that measurements of  current  $J_1$ into the ancilla bath can be used to determine the nature of the target bath, by detecting the value of $\mathcal{F}$ for a specific scenario}. To illustrate this point, a measurement protocol can be proposed:
(i)  $\Omega$ and $\Gamma_{1}$ are obtained from a measurement of the decay of the oscillator only coupled to the ancilla bath and $D=0$.
(ii)  $\Gamma_{2}$ is obtained from a similar measurement as before when the oscillator is coupled to both baths.
(iii) Measurements on the oscillator when $D=0$ give $E_{0}$ and $J_{1}^{(0)}$.
(iv) Energy or ancilla current measurements are performed for different values of $D$ (thus $\mathcal{W}$), obtaining from the slope the value of $\mathcal{F}$ (see also a thermometry protocol in App.\ref{app:Thermometry}).

Fig.~\ref{FigCombined}(c) shows the variation $\Delta \mathcal{F}_2\equiv |\mathcal{F}_{2,\rm Q}-\mathcal{F}_{2,\rm cl}|$ defined as the difference  in $\mathcal{F}$ when the ancilla bath is set to be quantum but the target is either quantum or classical, as a function of the fractional ancilla damping parameter $\gamma\equiv\Gamma_{1}/(\Gamma_{1}+\Gamma_{2})$ and the temperature ratio $T_2/T_1$ (log scale). We set $Q=10$, $Q_{\rm C}=10^{3}$ and $\widetilde{T}_{1}=1/4$.  The 12\% deviation in Fig. \ref{FigCombined}(b) stands as the maximum difference, occurring for small $\gamma$ (i.e., negligible contributions of the ancilla bath) and low temperatures of the target bath. As thermal differences increases, one of the two baths dominates over the other. If the target bath is the coldest, the differences are larger. Considering the dependence with $\gamma$, the difference increases when the target bath is dominant. However, decreasing $\gamma$ implies a smaller $J_{1}$, suggesting an interplay between having a dominant target bath and a measurable $J_{1}$. Differences up to 2\% are found over broad non-equilibrium conditions when the ancilla bath has the smallest damping rate.

{\it Conclusion}: We have shown, for the first time, that a stationary Brownian oscillator subject to frequency (multiplicative) noise can be used to assess the quantum or classical nature of a target bath, based on deviations from the classical virial ratio $\langle K\rangle/\langle V\rangle=1$, which can be magnified by a suitabe tuning of the noise strength $D$. This is achieved by either measuring the total oscillator energy $E$ or the heat current flowing to an ancilla bath, but without accessing or intervening directly the target bath. We show that due to stability constraints, lower quality oscillators ($Q<10$) subject to strong frequency driving are best suited for detecting deviations from the virial theorem, which are shown to exceed 10\% for thermal energies below the zero point motion of the driven oscillator ($2k_BT<\Omega$). Since frequency noise admits potential implementations in a variety of available physical platforms such as levitated nanoparticles in fluctuating optical tweezers \cite{LevitatedNanoparticleNovotny}, trapped ions in fluctuating Paul traps \cite{BlattReview}, optical cavities with fluctuating walls (see Appendix \ref{app:MultiplicativeNoiseImplementation}), and single molecules undergoing spectral diffusion \cite{Sarkar2024}, our theoretical predictions can be readily tested.

Generalizations of our results to oscillators with different spectral densities for the baths or coloured frequency noise are possible. Different physical implementations could be relevant in hybrid quantum platforms such as an ancilla bath being electromagnetic and the target bath being gravitational \citep{BeiLokGrav}, phononic, or plasmonic \cite{Shegai2019}, although case by case studies are required depending on the implementation. Our work thus opens avenues for sensing the fundamental nature of an environment, with perspectives towards the development of sensitive quantum thermometry.

\begin{acknowledgments}
We would like to thank Ricardo Decca, Felipe Recabal and Johannes Schachenmayer for helpful comments. A.E.R.L. and F.H. are supported by ANID through grants FONDECYT Iniciaci\'on No. 11250638, FONDECYT Regular No. 1221420 and the Millennium Science Initiative Program ICN17\_012.

\end{acknowledgments}

\bibliography{references-submit}

\newpage
\onecolumngrid

\clearpage

\appendix

\section{Frequency noise source for a specific implementation}\label{app:MultiplicativeNoiseImplementation}

This section is devoted to show the connection between the random variable $\varphi(t)$ and its correlation for a given implementation through the variation of a specific parameter. The analysis can be systematized for every harmonic physical system. Let us say that a physical system sets harmonic motion characterized by a frequency $\tilde{\Omega}^{2}(A)$ in such a way that it depends on a parameter $A$. This parameter is allowed to fluctuate in such a way that $A(t)=A_{0}+\delta A(t)$, where $A_{0}$ is the average value $\langle A(t)\rangle=A_{0}$ and $|\delta A|\ll A_{0}$. Then, we have:
\begin{equation}
\tilde{\Omega}^{2}(A_{0}+\delta A)\simeq\Omega^{2}+2\Omega\Omega'\delta A(t)+...=\Omega^{2}\left(1+2\frac{\Omega'}{\Omega}\delta A(t)+...\right),
\end{equation}
where $\Omega\equiv\tilde{\Omega}(A_{0})$ and $\Omega'\equiv\tilde{\Omega}'(A_{0})$.

In a similar way, by taking as a variation with respect to the the reference value $\tilde{\Omega}(A_{0}+\delta A)=\Omega+\delta\Omega$:
\begin{equation}
\tilde{\Omega}^{2}(A_{0}+\delta A)=(\Omega+\delta\Omega)^{2}=\Omega^{2}+2\Omega\delta\Omega+\delta\Omega^{2}\approx\Omega^{2}\left(1+2\frac{\delta\Omega}{\Omega}+\frac{\delta\Omega^{2}}{\Omega^{2}}\right)\approx\Omega^{2}\left(1+2\frac{\delta\Omega}{\Omega}\right).
\end{equation}

This gives two alternative expressions for the frequency noise:
\begin{equation}
\varphi(t)=2\frac{\Omega'\delta A(t)}{\Omega}=2\frac{\delta\Omega}{\Omega}.
\end{equation}

This way, the correlations of the random variable associated to the frequency noise is given by the multiple forms:
\begin{equation}
\left\langle\varphi(t)\varphi(t')\right\rangle=4\left\langle\frac{\delta\Omega(t)}{\Omega}\frac{\delta\Omega(t')}{\Omega}\right\rangle=4\left(\frac{\Omega'}{\Omega}\right)^{2}\left\langle\delta A(t)\delta A(t')\right\rangle.
\end{equation}

In this point we can define a spectral power density $S_{A}(f)$ associated to the fluctuations of the parameter $A$ having that:
\begin{equation}
\frac{1}{A_{0}^{2}}\left\langle\delta A(t)\delta A(t')\right\rangle=\frac{1}{2}\int_{-\infty}^{+\infty}df~S_{A}(f)~e^{-i2\pi f(t-t')},
\end{equation}
in such a way that the correlations of the multiplicative noise, in principle, reads:
\begin{equation}
\left\langle\varphi(t)\varphi(t')\right\rangle=2\left[\frac{\Omega'}{\Omega}A_{0}\right]^{2}\int_{-\infty}^{+\infty}df~S_{A}(f)~e^{-i2\pi f(t-t')}.
\end{equation}

The white noise limit is immediate by setting $S_{A}(f)\equiv\mathcal{S}_{A}$, obtaining:
\begin{equation}
\left\langle\varphi(t)\varphi(t')\right\rangle=2\left[\frac{\Omega'}{\Omega}A_{0}\right]^{2}\mathcal{S}_{A}~\delta(t-t'),
\end{equation}
where the latter immediately allow us to define $D=\mathcal{S}_{A}[\Omega'A_{0}/\Omega]^2$. For the particular case of a linear dependence of $\tilde{\Omega}^{2}(A)=\mathcal{C}A$, then $2\tilde{\Omega}(A)\tilde{\Omega}'(A)=\mathcal{C}$, in such a way that $\tilde{\Omega}'(A)=(1/2)\sqrt{\mathcal{C}/A}=\mathcal{C}/[2\tilde{\Omega}(A)]$, so finally we get $D=\mathcal{S}_{A}/4$.

In any case, we assume there is a minimum intrinsic level of noise, characterized by a minimal value of the power spectral density $\mathcal{S}_{\rm min}$. This value is physically expected by intrinsic white noises, such as Johnson noise in electronics or other varieties of shot noise. Given this value, we can define other noise levels in terms of decibels over it, according to the definition:
\begin{equation}
L_{A}\equiv 10~{\rm log}_{10}(\mathcal{S}_{A}/\mathcal{S}_{\rm min}),
\end{equation}
in such a way that we naturally have $\mathcal{S}_{A}=10^{\frac{L_{A}}{10}}\mathcal{S}_{\rm min}$. According to the definition of minimum value, $L_{A}\geq 0$. 

Following this train of thought, we can also define an enhancement factor $\Phi_{A}$ associated to the particular implementation of the multiplicative noise by taking the linear dependence case ($\tilde{\Omega}^{2}(A)=\mathcal{C}A$) as the reference:
\begin{equation}
\Phi_{A}\equiv 20~{\rm log}_{10}\left[2\frac{\Omega'}{\Omega}A_{0}\right],
\end{equation}
in such a way that the strength of the multiplicative noise correlations can be written:
\begin{equation}
D=10^{\frac{\Phi_{A}+L_{A}}{10}}\frac{\mathcal{S}_{\rm min}}{4},
\end{equation}
having that $\Phi_{A}$ and $L_{A}$ are measured in decibels. Notice that, for a given realization, the strength has its minimum value, in principle, at $D_{\rm min}\equiv10^{\frac{\Phi_{A}}{10}}\mathcal{S}_{\rm min}/4$.

Finally, we can obtain the correlations of the frequency variations as a white noise:
\begin{equation}
\left\langle\frac{\delta\Omega(t)}{\Omega}\frac{\delta\Omega(t')}{\Omega}\right\rangle=\frac{D}{2}~\delta(t-t').
\end{equation}

Given this general approach, we can apply it to specific implementations, namely, a levitated nanoparticle with a power-fluctuating optical tweezers, a trapped ion in a voltage-fluctuating Paul trap, and a cavity mode with a fluctuating wall.

\subsection{Levitating nanoparticle with power-fluctuating optical tweezers}\label{subapp:Nanoparticle}

As a first implementation, we consider a nanoparticle in an optical harmonic trap held by optical tweezers. In this case, the degree of freedom corresponds to the center of mass dynamics of the nanoparticle. Within this picture, the frequency characterizing the harmonic motion in each direction is given by the optical tweezers theory. The confined motion in trapping potentials given by optical tweezers is described in cylindrical coordinates. Thus, there are two possible frequencies characterizing the motion along the axial $z-$direction and the radial $r-$direction. We start just by considering the axial case, having that the frequency reads:
\begin{equation}
\Omega_{z}^{2}=\frac{2\alpha\lambda^{2}P}{M\pi^{3}\varepsilon_{0} cw_{0}^{6}},
\end{equation}
where $\alpha$ is the nanoparticle's polarizability, $\lambda$ the operation wavelength of the optical tweezers, $w_{0}$ stands for the minimum waist of the optical tweezers and $P$ corresponds to the power of the beam.

If we assume that the power of the beam can fluctuate, in such a way that $P(t)=P_{0}+\delta P(t)$. This way, we can write:
\begin{equation}
\Omega_{z}^{2}=\frac{2\alpha\lambda^{2}P_{0}}{M\pi^{3}\varepsilon_{0} cw_{0}^{6}}\left(1+\frac{\delta P(t)}{P_{0}}\right).
\end{equation}

In this model, we have $\varphi(t)=\delta P(t)/P_{0}$. Then, the correlations of the multiplicative noise directly relates to the power fluctuations of the beam, $\langle\varphi(t)\varphi(t')\rangle=\langle\delta P(t)\delta P(t')\rangle/P_{0}^{2}$. In the general case, we can write the power fluctuations' correlations in terms of the power spectral density $S_{P}(f)$:
\begin{equation}
\left\langle\varphi(t)\varphi(t')\right\rangle=\left\langle\frac{\delta P(t)}{P_{0}}\frac{\delta P(t')}{P_{0}}\right\rangle=\frac{1}{2}\int_{-\infty}^{+\infty}df~S_{P}(f)~e^{-i2\pi f(t-t')}\approx\frac{\mathcal{S}_{\rm P}}{2}\delta(t-t'),
\end{equation}
where the last equality corresponds to the white noise approximation.

If we consider that the correlation of fluctuations of the power is given by the shot noise affecting the trapping laser, then $\langle\delta P(t)\delta P(t')\rangle=h\nu P_{0}\delta(t-t')$. This way, shot noise is given as a white noise that allow us to write the correlations of the multiplicative noise as $\langle\varphi(t)\varphi(t')\rangle=(h\nu/P_{0})\delta(t-t')$, immediately giving $D_{\rm min}=h\nu/(2P_{0})$ for the optical tweezers model. For optical tweezers  with a power $P_{0}=0.5{\rm W}$ and a wavelength $\lambda=c/\nu=1.55\mu{\rm m}$, we have:
\begin{equation}
D_{\rm min}=\frac{hc}{2\lambda P_{0}}
=1.282...\times 10^{-19}{\rm s}.
\end{equation}

\subsection{Trapped ion in voltage-fluctuating Paul trap}\label{subapp:PaulTrap}

Another scenario can be the harmonic dynamics' of a trapped ion in a Paul trap. In this case, an oscillating quadrupole electric field is used for confining charges particles by circumventing Earnshaw's theorem for static fields.

The dynamics of an ion in a Paul trap is basically described by the Mathieu equation, reading in dimensionless form as:
\begin{equation}
\frac{d^{2}x}{d\tau^{2}}+\left[a-2q\cos\left(2\tau\right)\right]x(\tau)=0,
\end{equation}
where for a Paul trap $2\tau=\Omega t$, and the parameters are given by $a=[Q/(M\Omega^{2})]4V_{0}/r_{0}^{2}$ and $q=-[Q/(M\Omega^{2})]2V/r_{0}^{2}$. These definitions are attached to the applied voltage given by $V(t)=V_{0}+V\cos(\Omega t)$, while $Q$ ($M$) corresponds to the charge (mass) of the ion and $r_{0}$ is the distance characterizing the electrodes' spacing. Under some choices of parameters, the ion's motion consists of an oscillation at a slow secular frequency plus a fast, small-amplitude micromotion at the driving frequency in such a way that $x(t)\approx\mathcal{A}\cos(\omega t)[1+(q/2)\cos(\Omega t)]$. The slow motion of the ion is given by the secular frequency $\omega$, which under the Dehmelt approximation is (see Ref.\cite{BlattReview}):
\begin{equation}
\omega=\frac{\Omega}{2}\sqrt{a+\frac{q^{2}}{2}}.
\end{equation}

Given the latter, we can consider the impact of shot noise in dc component of the voltage ($V_{0}$). For this, we take $V_{0}\rightarrow V_{0}+\delta V$ and $a=AV_{0}$. Following the procedure at the beginning of the section, we obtain $\omega'=\Omega^{2}A/(8\omega)$. Now, taking the white noise limit for the voltage fluctuations $\langle\delta V(t)\delta V(t')\rangle/V_{0}^{2}=(\mathcal{S}_{V_{0}}/2)\delta(t-t')$. This immediately allow us to write the fluctuations of the multiplicative noise as:
\begin{eqnarray}
\left\langle\varphi(t)\varphi(t')\right\rangle&=&2\left(\frac{\Omega^{2}A}{8\omega^{2}}V_{0}\right)^{2}\mathcal{S}_{V_{0}}\delta(t-t')\\
&=&\frac{\mathcal{S}_{V_{0}}}{2\left(1+\frac{q^{2}}{2a}\right)^{2}}~\delta(t-t'),\nonumber
\end{eqnarray}
where $q^{2}/(2a)=|qV|/(4V_{0})$. This gives that $D=\mathcal{S}_{V_{0}}/(4[1+q^{2}/(2a)]^2)$. At first glance, notice that for $a=0$ (no dc component), there is no multiplicative noise.

Given that Paul traps consist in electrodes characterized by capacities ($C$), the voltage fluctuations translates into charge fluctuations according to $\langle\delta V(t)\delta V(t')\rangle=\langle\delta Q(t)\delta Q(t')\rangle/C^{2}$. Single-electron transistors shows that charge fluctuations are described by a white noise power spectral density $\mathcal{S}_{Q}\approx\hbar C$. This immediately gives that $\mathcal{S}_{V_{0}}\rightarrow\mathcal{S}_{\rm min}\approx\hbar/(CV_{0}^{2})$. A minimal value for the strength of the multiplicative noise is defined as $D_{\rm min}=10^{\frac{\Phi_{V_{0}}}{10}}\mathcal{S}_{\rm min}/4$, with:
\begin{equation}
\Phi_{V_{0}}=-20~{\rm log}_{10}\left|1+\frac{q^{2}}{2a}\right|.
\end{equation}

Typical numbers for a Paul trap are around a capacitance $C=10{\rm pF}$ and $|V_{0}|\sim(0-50)V$. As an extreme example, we take $V_{0}=-0.1{\rm V}$  we have a minimal value:
\begin{equation}
\mathcal{S}_{\rm min}=\frac{\hbar}{CV_{0}^{2}}
=1.054...\times 10^{-21}{\rm s},
\end{equation}
which in principle results a worst value than the optical tweezers case. However, in this case we have the decibels provided by the conditions under which the trap is built. Thus, we can choose the trap to give a decibel level of $\Phi_{V_{0}}=60{\rm dB}$, which immediately gives $D_{\rm min}\approx 2.5...\times 10^{-18}{\rm s}$, which is one order of magnitude larger than the optical tweezers case.

A similar situation to the previous one would be to consider fluctuations on the ac component of the voltage. In this case, the fluctuations of the multiplicative noise are given by:
\begin{eqnarray}
\left\langle\varphi(t)\varphi(t')\right\rangle&=&2\left(\frac{\Omega^{2}q^{2}}{8\omega^{2}}\right)^{2}\mathcal{S}_{V}\delta(t-t')\\
&=&\frac{2\mathcal{S}_{V}}{\left(1+\frac{2a}{q^{2}}\right)^{2}}~\delta(t-t'),\nonumber
\end{eqnarray}
which gives decibels levels:
\begin{equation}
\Phi_{V}=-20~{\rm log}_{10}\left|1+\frac{2a}{q^{2}}\right|.
\end{equation}

Now, suppose that we consider a situation where $q^{2}/(2a)\sim-1+\epsilon$, where $\epsilon\ll 1$. Then, $1+q^{2}/(2a)\sim\epsilon$. This implies that $1+2a/q^{2}\sim-\epsilon$. This is implying that in this scenario $\Phi_{V_{0}}\approx\Phi_{V}$. Thus, in the previous example where $\Phi_{V_{0}}=60{\rm dB}$, immediately we get $\Phi_{V}\approx 60{\rm dB}$. For a given realization of a Paul trap satisfying $q^{2}/(2a)\sim-1+\epsilon$, the noises in both components are similar.

\subsection{Cavity mode with fluctuating wall}\label{subapp:Cavity}

In this case we consider the possibility of frequency noise generated by the position fluctuations of one of the walls of a cavity. The degree of freedom then corresponds to the selected cavity mode taking part in the interaction. This mode is characterized by a wavelength $\lambda$. The jiggling wall's dynamics is described by its motion $z(t)$, which is a fluctuating stochastic variable which average $\langle z(t)\rangle=0$. We assume the statistics of the motion to be Gaussian, in such a way that the dynamics is fully characterized by the second order correlation of $z$, i.e., by $\langle z(t)z(t')\rangle$. Considering this, while the average wavelength of the mode is $\lambda$, the instantaneous wavelength is $\lambda'=\lambda-z$ (the minus sign is taken by convention as the vanishing average of $z$). For an empty cavity, the frequency of the cavity mode is directly connected to its wavelength by $\Omega'=2\pi c/\lambda'$. Then, we reasonably assume the motion of the wall to be confined over values much smaller than the average wavelength ($z\ll\lambda$). Thus, we can consider the square of frequency approximated as:
\begin{equation}
\Omega'^{2}=\frac{4\pi^{2}c^{2}}{\left(\lambda+z\right)^{2}}\approx\Omega^{2}+2\frac{4\pi^{2}c^{2}}{\left(\lambda-z\right)^{3}}\Bigg|_{z=0}z+...=\Omega^{2}+\frac{\Omega^{3}}{\pi c}z+...=\Omega^{2}\left(1+\frac{\Omega z}{\pi c}+...\right),
\end{equation}
where we have taken $\Omega=2\pi c/\lambda$. Thus, by discarding high-order corrections, we can match the frequency noise in our previous approach with the correction due to the wall's motion:
\begin{equation}
\varphi(t)\equiv \frac{2}{\lambda}z(t)
.
\end{equation}

Now, the second order correlation reads:
\begin{equation}
\left\langle \varphi(t)\varphi(t')\right\rangle=\frac{4}{\lambda^{2}}
\left\langle z(t)z(t')\right\rangle.
\end{equation}

As we are considering white noise for the multiplicative noise, we require $\langle \varphi(t)\varphi(t')\rangle=2D\delta(t-t')$. However, typical dissipative dynamics for degrees of freedom does not give delta-correlated motion. Thus, this path does not seem to be appropriate in this case.

Nevertheless, for a Brownian particle subjected to harmonic potential and under the influence of a quantum (Ohmic) thermal bath with cutoff-frequency function $f(\omega/\omega_{\rm C})$, we have that the correlations read:
\begin{eqnarray}
2\frac{\left\langle z(t)z(t')\right\rangle}{\lambda^{2}}&\rightarrow&\frac{4\hbar\gamma}{m\lambda^{2}}
\int_{0}^{+\infty}\frac{d\omega}{\pi}\frac{\omega\coth\left(\frac{\hbar
\omega}{2k_{\rm B}T}\right)f\left(\frac{\omega}{\omega_{\rm C}}\right)}{\left(\left[\omega^{2}-\omega_{0}^{2}\right]^{2}+4\gamma^{2}\omega^{2}\right)}\cos\left[\omega(t-t')\right]\\
&=&\frac{4\hbar\gamma}{m\lambda^{2}}
\int_{-\infty}^{+\infty}\frac{d\omega}{2\pi}\frac{\omega\coth\left(\frac{\hbar
\omega}{2k_{\rm B}T}\right)f\left(\frac{\omega}{\omega_{\rm C}}\right)}{\left(\left[\omega^{2}-\omega_{0}^{2}\right]^{2}+4\gamma^{2}\omega^{2}\right)}e^{-i\omega(t-t')}\nonumber\\
&=&\frac{1}{2}
\int_{-\infty}^{+\infty}\frac{d\omega}{2\pi}S_{z}(\omega)~e^{-i\omega(t-t')}.\nonumber
\end{eqnarray}
with:
\begin{equation}
S_{z}(\omega)=\frac{8\hbar\gamma}{m\lambda^{2}}\frac{\omega\coth\left(\frac{\hbar
\omega}{2k_{\rm B}T}\right)}{\left(\left[\omega^{2}-\omega_{0}^{2}\right]^{2}+4\gamma^{2}\omega^{2}\right)}f\left(\frac{\omega}{\omega_{\rm C}}\right).
\end{equation}

If we consider that $\omega_{\rm C}\gg\omega_{0}\gg\Omega$ (being $\Omega$ the system's frequency over which the cavity is acting as a multiplicative noise), then the power spectral density corresponds approximately to the value at $\omega=0$. This immediately implies that we can approximate the thermal factor by its argument, i.e., $\coth(\hbar\omega/[2k_{\rm B}T])\approx 2k_{\rm B}T/(\hbar\omega)$. Then, a white noise approximation for the multiplicative noise correlations is possible:
\begin{equation}
\left\langle \varphi(t)\varphi(t')\right\rangle\approx 2D\delta(t-t')~~,~~D=\frac{\mathcal{S}_{z}(0)}{2}=\frac{8\gamma k_{\rm B}T}{m\lambda^{2}\omega_{0}^{4}}.
\end{equation}

Notice that having that $\omega_{0}\gg\Omega$ and taking $\Omega=2\pi c/\lambda$, we immediately have $\lambda\omega_{0}\gg 2\pi c$. We can use this inequality to show that the product of the strength of the multiplicative noise and the frequency of the main oscillator is upper-bounded by the simpler expression:
\begin{equation}
D\Omega\ll\frac{1}{\pi^{2}}\frac{k_{\rm B}T}{mc^{2}}\frac{2\gamma}{\omega_{0}}\frac{\Omega}{\omega_{0}}.
\end{equation}

Estimating the order of magnitude of the upper bound gives an idea of the maximum strength possible within this white noise approximation. At first insight, it looks that each factor is expected to be quite small given the approximated regime considered. This gives an extremely low upper bound for the multiplicative noise's strength. Then,  a Brownian model generating white noise is not a suitable approach for an implementation.

\section{Dissipation and noise in the Caldeira-Leggett model}\label{app:CaldeiraLeggett}

In this section we summarize the approach of Brownian motion based on Caldeira-Leggett model for thermal baths linearly coupled to an oscillator. For this, we follow the approach given in Ref.\cite{BreuerPett}. Although it is a quantum approach for Heisenberg operators, equivalent procedures are valid in the classical. In that case, Heisenberg equations are replaced by classical Hamilton's equations of motion derived through the employment of Poisson brackets\cite{Goldstein}. All in all, we omit any operator notation.

We start by giving the Hamiltonians of the different parts of the total system, formed by a main oscillator linearly coupled to two thermal baths, given by a set of harmonic oscillators each:
\begin{equation}
H=K+V+\sum_{k=1}^{2}V_{c}(k)+\sum_{k=1}^{2}H_{\rm Int}(k)+\sum_{k=1}^{2}H_{B}(k),
\end{equation}
where $k$ labels the baths, while the kinetic and potential energies are given by:
\begin{equation}
K=\frac{p^{2}}{2M}~~,~~V=\frac{M\Omega^{2}}{2}x^{2},
\end{equation}
and the baths', interaction and counter-term Hamiltonians read:
\begin{equation}
H_{B}(k)=\sum_{n}\left(\frac{p_{kn}}{2m_{kn}}+\frac{m_{kn}\omega_{kn}^{2}}{2}q_{kn}^{2}\right)~~,~~H_{\rm Int}(k)=-x\sum_{n}\lambda_{kn}q_{kn}~~,~~V_{c}(k)=x^{2}\sum_{n}\frac{\lambda_{kn}^{2}}{2m_{kn}\omega_{kn}^{2}},
\end{equation}
where we are giving a definition for the stochastic fluctuation forces in terms of the dynamical variables of the oscillators of each bath and the coupling constants $\lambda_{kn}$.

Given the total Hamiltonian, the equations of motion are given by:
\begin{equation}
\dot{x}=\frac{p}{M}~~,~~\dot{q}_{kn}=\frac{p_{kn}}{m_{kn}},
\end{equation}
\begin{equation}
\dot{p}=-M\Omega^{2}x-x\sum_{k=1}^{2}\sum_{n}\frac{\lambda_{kn}^{2}}{m_{kn}\omega_{kn}^{2}}+\sum_{k=1}^{2}\sum_{n}\lambda_{kn}q_{kn}~~,~~\dot{p}_{kn}=-m_{kn}\omega_{kn}^{2}q_{kn}+\lambda_{kn}x.
\end{equation}

The first two equations can be used to obtain second order differential equations:
\begin{equation}
\ddot{x}+\Omega^{2}x+x\sum_{k=1}^{2}\sum_{n}\frac{\lambda_{kn}^{2}}{Mm_{kn}\omega_{kn}^{2}}-\frac{1}{M}\sum_{k=1}^{2}\sum_{n}\lambda_{kn}q_{kn}=0,
\label{EqMotionXCoupled}
\end{equation}
\begin{equation}
\ddot{q}_{kn}+\omega_{kn}^{2}q_{kn}-\frac{\lambda_{kn}}{m_{kn}}x=0.
\end{equation}

For solving the last equation, we define annihilation and creation operators for each bath mode $b_{kn},b_{kn}^{\dag}$ in such a way that the position and momentum operators at the initial time read $q_{kn}(0)=(b_{kn}+b_{kn}^{\dag})/\sqrt{2m_{kn}\omega_{kn}}$ and $p_{kn}(0)=-i\sqrt{m_{kn}\omega_{kn}/2}(b_{kn}-b_{kn}^{\dag})$, respectively. Then, the solution is given by:
\begin{equation}
q_{kn}(t)=\frac{1}{\sqrt{2m_{kn}\omega_{kn}}}\left(e^{-i\omega_{kn}t}b_{kn}+e^{i\omega_{kn}t}b_{kn}^{\dag}\right)+\frac{\lambda_{kn}}{m_{kn}\omega_{kn}}\int_{0}^{t}d\tau~\sin\left(\omega_{kn}(t-\tau)\right)x(\tau).
\end{equation}

Replacing this into Eq.(\ref{EqMotionXCoupled}), it gives:
\begin{equation}
\ddot{x}+\Omega^{2}x+x\sum_{k=1}^{2}\sum_{n}\frac{\lambda_{kn}^{2}}{Mm_{kn}\omega_{kn}^{2}}-\frac{1}{M}\int_{0}^{t}d\tau~D(t-\tau)x(\tau)=\frac{\xi(t)}{M},
\end{equation}
where the total dissipation kernel $D=D_{1}+D_{2}$ and the stochastic fluctuation force $\xi=\xi_{1}+\xi_{2}$ are given by:
\begin{equation}
D_{k}(t)=2\int_{0}^{+\infty}d\omega~\mathcal{J}_{k}(\omega)\sin\left(\omega t\right),
\end{equation}
\begin{equation}
\xi_{k}(t)=\sum_{n}\frac{\lambda_{kn}}{\sqrt{2m_{kn}\omega_{kn}}}\left(e^{-i\omega_{kn}t}b_{kn}+e^{i\omega_{kn}t}b_{kn}^{\dag}\right),
\end{equation}
where $\mathcal{J}_{k}(\omega)\equiv\sum_{n}(\lambda_{kn}^{2}/[m_{kn}\omega_{kn}])\delta(\omega-\omega_{kn})$ is the spectral density associated to the $k-$th bath, which then is replaced by a continuos function accounting for the irreversibility of the dissipative dynamics. In the Ohmic case, the replacement reads $\mathcal{J}_{k}(\omega)=(2M\Gamma_{k}\omega/\pi)f(\omega/\Omega_{\rm C})$, with the Lorentzian cutoff function $f(x)=1/(1+x^{2})$, being $\Gamma_{k}$ the damping constant of the corresponding bath.

Now, defining the damping kernel for each bath from the corresponding dissipation kernel as $d\tilde{\Gamma}_{k}/dt=-D_{k}(t)/M$, it is straightforward to prove that:
\begin{equation}
-\frac{1}{M}\int_{0}^{t}d\tau~D_{k}(t-\tau)x(\tau)=\frac{d}{dt}\left[\int_{0}^{t}d\tau~\tilde{\Gamma}_{k}(t-\tau)x(\tau)\right]-\frac{1}{M}\sum_{n}\frac{\lambda_{kn}^{2}}{m_{kn}\omega_{kn}^{2}},
\end{equation}
which in the replacement on the equation of motion, it cancels the potential counter-term, giving the well-known Langevin equation of the Brownian motion:
\begin{equation}
\ddot{x}+\Omega^{2}x+\frac{d}{dt}\left[\int_{0}^{t}d\tau~\tilde{\Gamma}(t-\tau)x(\tau)\right]=\frac{\xi(t)}{M},
\end{equation}
where similarly $\tilde{\Gamma}=\tilde{\Gamma}_{1}+\tilde{\Gamma}_{2}$.

Furthermore, the correlation of the stochastic fluctuation forces can be obtained by assuming a thermal state for each mode of a thermal bath, giving the noise kernels $N_{k}$ of Eq.(\ref{NoiseKernel}). On the other hand, the damping kernels read:
\begin{equation}
\tilde{\Gamma}_{k}(t)=4\Gamma_{k}\int_{0}^{+\infty}\frac{d\omega}{\pi}f\left(\frac{\omega}{\Omega_{\rm C}}\right)\cos\left(\omega t\right).
\end{equation}

At this point we observe that the large cutoff limit $\Omega_{\rm C}\gg\Omega$ implies that the damping kernel is peaked on time with respect to the typical dynamical timescale of the oscillator. Thus, $f(\omega/\Omega_{\rm C})\approx 1$ for the relevant time step in the evolution, immediately localizing the damping kernel in a very accurate approximation:
\begin{equation}
\tilde{\Gamma}_{k}(t)\approx 4\Gamma_{k}\delta(t).
\end{equation}

The latter allow us the simplification of the damping term in the equation of motion, giving the Langevin equation of Markovian type:
\begin{equation}
\ddot{x}+\Omega^{2}x+4\Gamma\dot{x}=\frac{\xi(t)}{M}.
\end{equation}

All in all, for the final equation of motion of Eq.(\ref{EqMotion}), it is required to include the multiplicative frequency noise described by $\varphi$.

\section{Approach for the energy of Brownian particle under multiplicative noise at the steady state}\label{app:WestApproachForEnergy}

This section is devoted to summarize the approach of Ref.\cite{West1980} for a Brownian oscillator under the influence of frequency noise and in a out of equilibrium scenario given by the interaction with two thermal baths. Then, we derive expressions the kinetic, potential and total energies of the oscillator in the steady state. Finally, we bring the total energy into the form of Eq.(\ref{EnergyWithFreqNoise}).

We start directly from Eq.(\ref{EqMotion}), describing a Brownian oscillator including a frequency noise $\varphi$ and two independent ohmic thermal baths, both of them in the combined regime of large cutoff:
\begin{equation}
\ddot{x}+\Omega^{2}\left[1+\varphi(t)\right]x+4\Gamma\dot{x}(t)=\frac{\xi(t)}{M},
\end{equation}
where the damping constant $\Gamma=\Gamma_{1}+\Gamma_{2}$, the stochastic source $\xi=\xi_{1}+\xi_{2}$ and the correlations $\langle\xi_{j}(t)\xi_{k}(\tau)\rangle=2\delta_{jk}\tilde{D}_{j}\Phi_{j}(t-\tau)$ are given in the form of Ref.\cite{West1980}, in such a way that $\langle\xi(t)\xi(\tau)\rangle=\langle\xi_{1}(t)\xi_{1}(\tau)\rangle+\langle\xi_{2}(t)\xi_{2}(\tau)\rangle=2\tilde{D}_{1}\Phi_{1}(t-\tau)+2\tilde{D}_{2}\Phi_{2}(t-\tau)$. In our case, we are interested in noises resulting from thermal (quantum or classical) baths given as a set of harmonic oscillators. Considering a quantum origin for the environments, the coefficients $\tilde{D}_{j}$ and the noise functions $\Phi_{j}$ are given by:
\begin{equation}
\tilde{D}_{j}=2M\Gamma_{j}k_{\rm B}T_{j}~~,~~\Phi_{j}(t-\tau)=\frac{1}{2k_{\rm B}T_{j}}\int_{0}^{+\infty}\frac{d\omega}{\pi}~\omega~f\left(\frac{\omega}{\Omega_{\rm C}}\right)
\coth\left(\frac{\omega}{2k_{\rm B}T_{j}}\right)\cos\left[\omega(t-\tau)\right],
\label{DjPhi}
\end{equation}
where $\Omega_{\rm C}$ is the baths' cutoff frequency, $f(x)=1/(1+x^{2})$ is the Lorentzian cutoff function, in such a way that the product gives the noise kernels of Eq.(\ref{NoiseKernel}).

On the other hand, the frequency noise is characterized as white noise, fundamentally given by its second moment $\langle\varphi(t)\varphi(t')\rangle=2D\delta(t-t')$, while in general the cumulants are assumed as:
\begin{equation}
\langle\langle\varphi(t_{1})...\varphi(t_{n})\rangle\rangle=2^{n}D_{n}\delta(t_{1}-t_{2})...\delta(t_{n-1}-t_{n}),
\label{CumulantsVarphi}
\end{equation}
with $D_{n}$ the corresponding strengths for the higher-order cumulants.

In Ref.\cite{West1980}, the solution for this scenario is obtained. For $\mathbf{Y}=(x,p)$, the solution is given by:
\begin{equation}
\mathbf{Y}=\mathbb{S}^{-1}\cdot\mathbf{A},
\label{DefinitionY}
\end{equation}
for which considering that $\tilde{\Omega}=\sqrt{\Omega^{2}-4\Gamma^{2}}=\Omega\sqrt{1-1/(2Q)^{2}}=2\Gamma\sqrt{4Q^{2}-1}$, where $Q\equiv\Omega/4\Gamma$ stands for the quality factor of the oscillator comparing the sum of the dampings with the natural frequency, $\lambda_{\pm}=-2\Gamma\pm i\tilde{\Omega}$ and $e^{i\theta}\equiv\lambda_{-}/\lambda_{+}$, we have:
\begin{equation}
\mathbb{S}^{-1}=\frac{\Omega^{2}}{2i\tilde{\Omega}}
\begin{pmatrix}
  -1/\lambda_{+} & 1/\lambda_{-} \\
  -1 & 1  \\
\end{pmatrix}~~,~~\mathbf{A}=\int_{0}^{t}dt_{1}~e^{\Lambda t}\cdot\mathbb{Q}(t,t_{1})\cdot\xi_{\rm T}(t_{1})\mathbf{g}(t_{1}),
\end{equation}
\begin{equation}
\Lambda=\begin{pmatrix}
  \lambda_{+} & 0 \\
  0 & \lambda_{-}  \\
\end{pmatrix}~~,~~\mathbb{Q}(t,t_{1})\equiv T\left({\rm exp}\left[-\Omega^{2}\int_{t_{1}}^{t}d\tau~\varphi(\tau)\tilde{\mathbb{B}}(\tau)\right]\right)~~,~~\mathbf{g}(t)=\begin{pmatrix}
  -e^{-\lambda_{+}t}/\lambda_{-} \\
  -e^{-\lambda_{-}t}/\lambda_{+}  \\
\end{pmatrix},
\label{LambdaQg}
\end{equation}
\begin{equation}
\tilde{\mathbb{B}}(t)=\frac{1}{2i\tilde{\Omega}}\begin{pmatrix}
  1 & -e^{-i(2\tilde{\Omega}t+\theta)} \\
  e^{i(2\tilde{\Omega}t+\theta)} & -1  \\
\end{pmatrix},
\end{equation}
where the latter is a nilpotent matrix, such that $\tilde{\mathbb{B}}^{2}(t)=0$. Exploiting this property together with the white noise condition of the frequency noise, it is possible to solve for the dynamics of the oscillator and calculate the first and second moments for the dynamical variables $\{x,p\}$. In Ref.\cite{West1980}, according to our notation, the authors showed that in the long-time limit:
\begin{equation}
\lim_{t\rightarrow+\infty}\left\langle x^{2}\right\rangle(t)=\frac{\left(\tilde{\mathcal{D}}_{c}+\tilde{\mathcal{D}}_{s}\right)}{M^{2}\Omega^{2}\left(1-QD\Omega\right)}~~~~,~~~~\lim_{t\rightarrow+\infty}\left\langle xp\right\rangle(t)=0~~,~~\lim_{t\rightarrow+\infty}\left\langle p^{2}\right\rangle(t)=\frac{\left(\tilde{\mathcal{D}}_{c}+\left[2QD\Omega-1\right]\tilde{\mathcal{D}}_{s}\right)}{\left(1-QD\Omega\right)},
\label{SecondMomentsExpectation}
\end{equation}
where we have:
\begin{equation}
\begin{Bmatrix}
  \tilde{\mathcal{D}}_{c} \\
  \tilde{\mathcal{D}}_{s}  \\
\end{Bmatrix}=\begin{Bmatrix}
  \tilde{\mathcal{D}}_{c}^{(1)}+\tilde{\mathcal{D}}_{c}^{(2)} \\
  \tilde{\mathcal{D}}_{s}^{(1)}+\tilde{\mathcal{D}}_{s}^{(2)}  \\
\end{Bmatrix}=\frac{1}{2\Gamma}\int_{0}^{+\infty}d\tau~\begin{Bmatrix}
  \cos\left(\tilde{\Omega}\tau\right) \\
  \sin\left(\tilde{\Omega}\tau\right)/\sqrt{4Q^{2}-1}  \\
\end{Bmatrix}e^{-2\Gamma\tau}\left[\tilde{D}_{1}\Phi_{1}(\tau)+\tilde{D}_{2}\Phi_{2}(\tau)\right].
\label{DcDsFormal}
\end{equation}

Using Eq.(\ref{DjPhi}), we immediately get:
\begin{equation}
\begin{Bmatrix}
  \tilde{\mathcal{D}}_{c}^{(k)} \\
  \tilde{\mathcal{D}}_{s}^{(k)}  \\
\end{Bmatrix}
=M\Gamma_{k}\int_{0}^{+\infty}\frac{du}{\pi}f\left(\frac{u}{u_{\rm C}}\right)
\begin{Bmatrix}
  (1+u^{2}) \\
  (1-u^{2})  \\
\end{Bmatrix}\frac{u}{\left([1-u^{2}]^{2}+\frac{u^{2}}{Q^{2}}\right)}\coth\left(\frac{u}{\widetilde{T}_{k}}\right),
\label{DckDsk}
\end{equation}
where we have defined $u_{\rm C}\equiv\Omega_{\rm C}/\Omega$ and a dimensionless thermal energy $\widetilde{T}_{k}\equiv 2k_{\rm B}T_{k}/\Omega$. It is worth mentioning that in the high-temperature limit, we have $\coth(u/\tilde{T}_{k})\approx\tilde{T}_{k}/u$ in such a way that $\tilde{\mathcal{D}}_{c}^{(k)}\approx\tilde{D}_{k}/(4\Gamma)$ while $\tilde{\mathcal{D}}_{s}^{(k)}\approx0$.

This way, we can obtain for the long-time limit of the kinetic and potential energies ($\langle K\rangle\equiv\lim_{t\rightarrow+\infty}\langle p^{2}\rangle/(2M)$  and $\langle V\rangle\equiv\lim_{t\rightarrow+\infty}M\Omega^{2}\langle x^{2}\rangle/2$ , respectively) that:
\begin{equation}
\langle K\rangle^{(D)}=\langle K\rangle+\frac{QD\Omega}{\left(1-QD\Omega\right)}\langle V\rangle~~~,~~~\langle V\rangle^{(D)}=\frac{\langle V\rangle}{\left(1-QD\Omega\right)},
\end{equation}
with the kinetic and potential energies without multiplicative noise ($D=0$) are the ones resulting from the Brownian motion theory, reading:
\begin{equation}
\begin{Bmatrix}
  \langle K\rangle \\
  \langle V\rangle  \\
\end{Bmatrix}
=\int_{0}^{+\infty}\frac{du}{\pi}
\frac{u}{\left([1-u^{2}]^{2}+\frac{u^{2}}{Q^{2}}\right)}\begin{Bmatrix}
  u^{2}f\left(\frac{u}{u_{\rm C}}\right) \\
  1  \\
\end{Bmatrix}\sum_{k}\Gamma_{k}\coth\left(\frac{u}{\widetilde{T}_{k}}\right).
\end{equation}

Notice that for the potential energy no cutoff function is required since the integral naturally converges for every temperature. On the other hand, the kinetic energy requires a cutoff function for converging and the value of the integral is heavily depending on the cutoff frequency contained in $u_{\rm C}$.

Going further, each thermal factor is separated in zero-point fluctuations plus thermal contributions, according to $\coth(u/\widetilde{T}_{k})=1+2\overline{n}_{k}(u)$, in such a way that:
\begin{equation}
\begin{Bmatrix}
  \langle K\rangle \\
  \langle V\rangle  \\
\end{Bmatrix}
=\frac{\Omega}{2}\int_{0}^{+\infty}\frac{du}{2\pi}
\frac{u}{Q\left([1-u^{2}]^{2}+\frac{u^{2}}{Q^{2}}\right)}\begin{Bmatrix}
  u^{2}f\left(\frac{u}{u_{\rm C}}\right) \\
  1  \\
\end{Bmatrix}\left(1+2\sum_{k}\gamma_{k}\overline{n}_{k}(u)\right),
\end{equation}
where $\gamma_{k}\equiv\Gamma_{k}/\Gamma$, with $\Gamma\equiv\sum_{m}\Gamma_{m}$.

For the single bath case, the last equations lead to the expressions in Eq.(\ref{KinAndPot}). For the case of two baths ($k=1,2$), we can set $\gamma\equiv\gamma_{1}$ while $\gamma_{2}=1-\gamma$, in such a way that:
\begin{equation}
\begin{Bmatrix}
  \langle K\rangle \\
  \langle V\rangle  \\
\end{Bmatrix}
=\frac{\Omega}{2}\int_{0}^{+\infty}\frac{du}{2\pi}
\frac{u}{Q\left([1-u^{2}]^{2}+\frac{u^{2}}{Q^{2}}\right)}\begin{Bmatrix}
  u^{2}f\left(\frac{u}{u_{\rm C}}\right) \\
  1  \\
\end{Bmatrix}\left(1+2\left[\gamma\overline{n}_{1}(u)+(1-\gamma)\overline{n}_{2}(u)\right]\right).
\label{KVDoubleOccupation}
\end{equation}

From these expressions we can obtain the energy for the oscillator in the steady state by:
\begin{equation}
E\equiv \langle K\rangle^{(D)}+\langle V\rangle^{(D)}=\langle K\rangle+\frac{\left(1+QD\Omega\right)}{\left(1-QD\Omega\right)}\langle V\rangle=\frac{\left(\tilde{\mathcal{D}}_{c}+QD\Omega~\tilde{\mathcal{D}}_{s}\right)}{M\left(1-QD\Omega\right)}.
\end{equation}

Immediately, we can notice that for $D=0$ (no fluctuations in frequency or multiplicative noise), we have $E\rightarrow E_{0}\equiv \langle K\rangle+\langle V\rangle=\tilde{\mathcal{D}}_{c}/M$ independently of the equilibrium or nonequilibrium conditions. Considering that $\lim_{t\rightarrow+\infty}\langle x\rangle(t)=0$ regardless on the value of $D$, we can define $\Delta x_{0}\equiv\Delta x|_{D=0}=\lim_{t\rightarrow+\infty}[\langle x^{2}(t)\rangle-\langle x(t)\rangle^{2}]|_{D=0}=\lim_{t\rightarrow+\infty}\langle x^{2}(t)\rangle|_{D=0}$, in such a way that that the energy in the steady state can be re-written as:
\begin{equation}
E=E_{0}+\frac{QD\Omega}{\left(1-QD\Omega\right)}\Delta E,
\end{equation}
where $\Delta E=M\Omega^{2}\Delta x_{0}$ is the energy excess due to a frequency driving characterized by $D$ on an damped oscillator with a quality factor $Q$. Notice that $\Delta E$ is independent of $D$ and just depends on the position fluctuations of the oscillator without the driving applied. Furthermore, we can write $\Delta E=2\langle V\rangle$. Then, we have $\Delta E/E_{0}=2/(1+\mathcal{R})$, with $\mathcal{R}\equiv \langle K\rangle/\langle V\rangle$ being the Virial ration, in such a way that:
\begin{equation}
E=E_{0}\left(1+\mathcal{W}
\mathcal{F}
\right).
\end{equation}
%
%
%
where the magnification factor is $\mathcal{W}=QD\Omega/(1-QD\Omega)$ and the Virial deviation factor reads $\mathcal{F}=2/(1+\mathcal{R})$, resulting in Eq.(\ref{EnergyWithFreqNoise}).

%

\section{Calculation of heat currents}\label{app:HeatCurrents}

This section is devoted to the extension of the approach Ref.\cite{West1980} to calculate the heat currents taking place in an oscillator interacting with two thermal baths and under the influence of frequency noise, and to finally show the expressions of Eq.(\ref{HeatCurrentsFull}). It is worth mentioning that the result of Eq.(3) and the equations obtained in the last section remain valid since they are completely general at the steady state of an oscillator under the influence of an arbitrary number of thermal ohmic reservoirs at arbitrary temperatures. However, at an out of equilibrium scenario, in addition to the energy, heat currents arise due to the thermal imbalance.

For obtaining general expressions for the heat currents, we go back to the equation of motion Eq.(\ref{EqMotion}) for a single oscillator with two independent additive noises ($\xi_{1,2}$) and one frequency noise ($\varphi$), which in the large cutoff regime involve two different damping rates ($\Gamma_{1,2}$):
\begin{equation}
\ddot{x}(t)+\Omega^{2}\left[1+\varphi(t)\right]x(t)+4\left(\Gamma_{1}+\Gamma_{2}\right)\dot{x}(t)=\frac{\xi_{1}(t)}{M}+\frac{\xi_{2}(t)}{M}.
\end{equation}

As a first step, we trivially split this second-order differential equation into two first-order ones by including the definition of momentum:
\begin{equation}
\dot{x}=\frac{p}{M}~~,~~\dot{p}+M\Omega^{2}\left[1+\varphi(t)\right]x+4\left(\Gamma_{1}+\Gamma_{2}\right)p=\xi_{1}+\xi_{2}.
\end{equation}

From these equations, we go directly to the second moments. As before, we start by writing the equations for the squares of each canonical variable from the main equations of motion:
\begin{equation}
\frac{d x^{2}}{dt}-\frac{2}{M}xp=0~~,~~\frac{dp^{2}}{dt}+8\left(\Gamma_{1}+\Gamma_{2}\right)p^{2}+2M\Omega^{2}xp+2M\Omega^{2}\varphi xp-2\left(\xi_{1}+\xi_{2}\right)p=0,
\label{P2DoubleBath}
\end{equation}
which for the statistical expectation value they immediately give:
\begin{equation}
\frac{d\langle x^{2}\rangle}{dt}-\frac{2}{M}\langle xp\rangle=0~~,~~\left(\frac{d}{dt}+8\Gamma_{1}+8\Gamma_{2}\right)\langle p^{2}\rangle+2M\Omega^{2}\langle xp\rangle+2M\Omega^{2}\langle\varphi xp\rangle-2\langle\xi_{1}p\rangle-2\langle\xi_{2}p\rangle=0.
\label{X2P2Time}
\end{equation}

For analyzing the heat transfer we calculate the variation of the energy of the oscillator with respect to time, according to Ref.\cite{HeatTransferQBM}. This is straightforward from the definition of the energy of the oscillator and exploying the last averaged equations:
\begin{equation}
\frac{dE}{dt}\equiv\frac{d}{dt}\left[\frac{\left\langle p^{2}\right\rangle}{2M}+\frac{M\Omega^{2}}{2}\left\langle x^{2}\right\rangle\right]=-\frac{4}{M}\left(\Gamma_{1}+\Gamma_{2}\right)\left\langle p^{2}\right\rangle-\Omega^{2}\left\langle\varphi xp\right\rangle+\frac{\left\langle\xi_{1}p\right\rangle}{M}+\frac{\left\langle\xi_{2}p\right\rangle}{M}.
\end{equation}

In the steady state, we have that $dE/dt=0$, which immediately gives us a balance equation between the three sources of energy taking place in the system, namely, each thermal bath (additive noises) and the noise in the frequency:
\begin{equation}
\left\langle\xi_{1}p\right\rangle+\left\langle\xi_{2}p\right\rangle-4\left(\Gamma_{1}+\Gamma_{2}\right)\left\langle p^{2}\right\rangle-M\Omega^{2}\left\langle\varphi xp\right\rangle=0.
\end{equation}

Notice that for no frequency noise ($\varphi=0$), we re-obtain the expression considered in Ref.\cite{HeatTransferQBM}. Despite on the contribution of the frequency noise, the energy flowing to each of the thermal baths can be associated to the heat current:
\begin{equation}
J_{k}\equiv-\frac{4\Gamma_{k}}{M}\left\langle p^{2}\right\rangle+\frac{\left\langle\xi_{k}p\right\rangle}{M},
\label{HeatCurrentDefinition}
\end{equation}
in such a way that the steady state equation can be re-written as:
\begin{equation}
J_{1}+J_{2}=W,
\end{equation}
where we have defined the work power provided by the multiplicative noise as the remaining term $W\equiv\Omega^{2}\left\langle\varphi xp\right\rangle$.

Given these definitions, we proceed to the calculation of the currents by extending the method developed Ref.\cite{West1980} to the required expectation values. For the heat currents flowing through the system we require the expectation values of the products $\langle\xi_{k}p\rangle$. Employing the definition given in Eq.(\ref{DefinitionY}), we have:
\begin{eqnarray}
\left\langle\xi_{k}(t)\mathbf{Y}(t)\right\rangle
&=&\int_{0}^{t}dt_{1}~\mathbb{S}^{-1}\cdot e^{\Lambda t}\cdot\left\langle\mathbb{Q}(t,t_{1})\right\rangle\cdot\left\langle\xi_{k}(t)\xi_{\rm T}(t_{1})\right\rangle\mathbf{g}(t_{1}).
\label{XikY}
\end{eqnarray}

Given the definition in Eq.(\ref{LambdaQg}) and the identity found in Ref.\cite{Kubo1963}, it is possible to write the expectation value of $\mathbb{Q}$ in terms of the cumulants of the frequency noise variable $\varphi$, having:
\begin{eqnarray}
\left\langle\mathbb{Q}(t,t_{1})\right\rangle&=&\left\langle T\left({\rm exp}\left[-\Omega^{2}\int_{0}^{t}d\tau~\varphi(\tau)\tilde{\mathbb{B}}(\tau)\right]\right)\right\rangle\\
&=&{\rm exp}\left[\sum_{n=1}^{+\infty}(-1)^{n}\Omega^{2n}\int_{0}^{t}d\tau_{1}...\int_{0}^{t_{n}}d\tau_{n}\langle\langle\varphi(\tau_{1})...\varphi(\tau_{n})\rangle\rangle\tilde{\mathbb{B}}(\tau_{1})...\tilde{\mathbb{B}}(\tau_{n})\right].\nonumber
\end{eqnarray}
%

Now, considering that the cumulants of the frequency noise are given by Eq.(\ref{CumulantsVarphi}), we obtain:
\begin{equation}
\left\langle\mathbb{Q}(t,t_{1})\right\rangle={\rm exp}\left[\sum_{n=1}^{+\infty}(-2)^{n}\Omega^{2n}D_{n}\int_{0}^{t}d\tau_{1}...\int_{0}^{t_{n}}d\tau_{n}\delta(\tau_{1}-\tau_{2})...\delta(\tau_{n-1}-\tau_{n})\tilde{\mathbb{B}}(\tau_{1})...\tilde{\mathbb{B}}(\tau_{n})\right].
\end{equation}
Thus, having that $\langle\varphi(t)\rangle=0$ by definition, the first term of the sum vanishes. Then, for the terms with $n\geq 2$, the Dirac deltas combined with the nilpotency of $\tilde{\mathbb{B}}(t)$, implies that all the other terms vanish too. Finally, we obtain that $\langle\mathbb{Q}(t,t_{1})\rangle=\mathbb{I}$.

Going back to Eq.(\ref{XikY}), and considering the independence of the additive noises ($\langle\xi_{1}\xi_{2}\rangle=0$), we immediately have:
\begin{eqnarray}
\left\langle\xi_{k}(t)\mathbf{Y}(t)\right\rangle&=&\frac{\tilde{D}_{k}\Omega^{2}}{i\tilde{\Omega}}\int_{0}^{t}dt_{1}~\Phi_{k}(t-t_{1})
\begin{pmatrix}
  -1/\lambda_{+} & 1/\lambda_{-} \\
  -1 & 1  \\
\end{pmatrix}\cdot \begin{pmatrix}
  -e^{\lambda_{+}(t-t_{1})}/\lambda_{-} \\
  -e^{\lambda_{-}(t-t_{1})}/\lambda_{+}  \\
\end{pmatrix}
.
\end{eqnarray}

Thus, the desired expectation values read:
\begin{eqnarray}
\left\langle\xi_{k}(t)p(t)\right\rangle
&=&\frac{2}{\tilde{\Omega}}\int_{0}^{t}d\tau~{\rm Im}\left[\lambda_{+}e^{i\tilde{\Omega}\tau}\right]e^{-2\Gamma_{\rm T}\tau}\tilde{D}_{k}\Phi_{k}(\tau)\nonumber\\
&=&2\int_{0}^{t}d\tau~\left[\cos\left(\tilde{\Omega}\tau\right)-\frac{\sin\left(\tilde{\Omega}\tau\right)}{\sqrt{4Q^{2}-1}}\right]e^{-2\Gamma_{\rm T}\tau}\tilde{D}_{k}\Phi_{k}(\tau),\nonumber
\end{eqnarray}
in such a way that the long-time limit immediately reads:
\begin{eqnarray}
\lim_{t\rightarrow+\infty}\left\langle\xi_{k}(t)p(t)\right\rangle&=&2\int_{0}^{+\infty}d\tau~\left[\cos\left(\tilde{\Omega}\tau\right)-\frac{\sin\left(\tilde{\Omega}\tau\right)}{\sqrt{4Q^{2}-1}}\right]e^{-2\Gamma_{\rm T}\tau}\tilde{D}_{k}\Phi_{k}(\tau)\\
&=&4\Gamma_{\rm T}\left(\tilde{\mathcal{D}}_{c}^{(k)}-\tilde{\mathcal{D}}_{s}^{(k)}\right).\nonumber
\end{eqnarray}

Having that the heat currents are given by Eq.(\ref{HeatCurrentDefinition}), we obtain:
\begin{equation}
J_{k}=-\frac{4\Gamma_{k}}{M}\frac{\left(\tilde{\mathcal{D}}_{c}+\left[2QD\Omega-1\right]\tilde{\mathcal{D}}_{s}\right)}{\left(1-QD\Omega\right)}+\frac{4\Gamma_{\rm T}}{M}\left(\tilde{\mathcal{D}}_{c}^{(k)}-\tilde{\mathcal{D}}_{s}^{(k)}\right).
\end{equation}

By defining the current for $D=0$, we have:
\begin{equation}
J_{k}^{(0)}=-\frac{4\Gamma_{k}}{M}\left(\tilde{\mathcal{D}}_{c}-\tilde{\mathcal{D}}_{s}\right)+\frac{4\Gamma_{\rm T}}{M}\left(\tilde{\mathcal{D}}_{c}^{(k)}-\tilde{\mathcal{D}}_{s}^{(k)}\right).
\end{equation}

Notice that we immediately get:
\begin{equation}
J_{1}^{(0)}=\frac{4\Gamma_{2}}{M}\left(\tilde{\mathcal{D}}_{c}^{(1)}-\tilde{\mathcal{D}}_{s}^{(1)}\right)-\frac{4\Gamma_{1}}{M}\left(\tilde{\mathcal{D}}_{c}^{(2)}-\tilde{\mathcal{D}}_{s}^{(2)}\right)=-J_{2}^{(0)},
\end{equation}
which is giving that the current flowing from one of the reservoirs is the same as the one flowing from the other. Given the energy conservation of the system, the two currents complement each other. They are symmetric, as physically expected. This symmetry gets broken when the frequency is stochastically driven.

Now, for the general case, when frequency driving is applied, combining Eqs.(\ref{DcDsFormal}) and (\ref{DckDsk}) we get:
\begin{equation}
J_{k}=
J_{k}^{(0)}-4\Gamma_{k}E_{0}\mathcal{W}\mathcal{F},
\end{equation}
which shows Eq.(\ref{HeatCurrentsFull}).

As a last comment, notice that as it is written, the modification due to the frequency driving always substract to the current at zero multiplicative noise. As for the energy case, we can proceed as before an exploit the Virial Theorem again by using that $J_{1}^{(0)}+J_{2}^{(0)}=0$, in such a way that:
\begin{equation}
-\frac{(J_{1}+J_{2})}{4\Gamma_{\rm T}E_{0}}=\frac{QD\Omega}{(1-QD\Omega)}\frac{2}{(1+\mathcal{R})},
\end{equation}
which seems to be a fundamental relation as it happens for $E/E_{0}-1$.

\section{Application: Thermometry scheme}\label{app:Thermometry}

Given the energy and the currents of Eqs.(\ref{EnergyWithFreqNoise}) and (\ref{HeatCurrentsFull}), we employ the results to study the advantages that a scheme based on frequency noise provides.

A first example consists in setting a scheme for thermometry based on the stochastic frequency oscillator interacting with two thermal baths at different temperatures. In this case, we consider $T_{1}$ as a reference temperature set in the ancilla bath, while the target bath is at temperature $T_{2}=T_{1}+\Delta T$. The difference $\Delta T$ is, in principle, arbitrary. Additionally, we assume that both baths are Ohmic and their nature (classical or quantum) is known.

Now, for the derivation, it is useful to identify the dependences on the temperatures by each of the quantities. While $\mathcal{W}$ is temperature independent, the virial factor is temperature dependent $\mathcal{F}=\mathcal{F}(T_{1},T_{1}+\Delta T)$. Then, similarly happens for the energy $E=E(T_{1},T_{1}+\Delta T)$ and the currents $J_{k}=J_{k}(T_{1},T_{1}+\Delta T)$, and also when the frequency noise is not applied, i.e., for $E_{0}$ and $J_{k}^{(0)}$.

Given this situation, we propose measurement protocols for thermometry of the target bath.

For large thermal differences ($\Delta T\sim T_{1}$), either the energy or the current are expected present large variations, independently on the fact that the frequency noise is applied or not. However, when the difference are small ($\Delta T\ll T_{1}$) and no frequency noise is applied ($D=0$), the precision of thermometers might reach a limitation in distinguishing two close temperature values because of the small variations that it generates such difference in the energy $E_{0}$ or the currents $J_{k}^{(0)}$ with respect to the equilibrium values, namely $E_{0}^{\rm (Eq)}$ and, by definition, a vanishing current $J_{k}^{(0),\rm Eq}=0$. This means that:
\begin{equation}
E_{0}(T_{1},T_{1}+\Delta T;Q)\approx E_{0}^{\rm (Eq)}(T_{1};Q)+C_{0}\Delta T+...~~~,~~~J_{k}^{(0)}(T_{1},T_{1}+\Delta T)\approx K_{0}\Delta T+....
\end{equation}
having defined $C_{0}\equiv\left.\frac{\partial E_{0}}{\partial\Delta T}\right|_{\Delta T=0}$ and $K_{0}\equiv\left.\frac{\partial J_{k}^{(0)}}{\partial\Delta T}\right|_{\Delta T=0}$. As we said, it is expected that both corrections are found to be small for $\Delta T\ll T_{1}$, such that $C_{0}\Delta T\ll E_{0}^{\rm (Eq)}$ and $K_{0}\Delta T\approx 0$.

However, when the frequency noise is applied, the virial factor is also expected to present a variation such that:
\begin{equation}
\mathcal{F}(T_{1},T_{1}+\Delta T;Q)\approx\mathcal{F}^{\rm (Eq)}(T_{1};Q)\left(1-\frac{\mathcal{F}^{\rm (Eq)}(T_{1};Q)}{2}\left.\frac{\partial}{\partial\Delta T}\Big[\mathcal{R}(T_{1},T_{1}+\Delta T;Q)\Big]\right|_{\Delta T=0}\Delta T+...\right).
\end{equation}

Then, controlling the amplification factor such that we can be made arbitrarily large ($\mathcal{W}\gg 1$), so the main the contribution is given by that term, we have:
\begin{equation}
E(T_{1},T_{1}+\Delta T;Q)
\approx\mathcal{W}E_{0}^{\rm (Eq)}(T_{1};Q)\mathcal{F}^{\rm (Eq)}(T_{1};Q)\left[1-\frac{\mathcal{F}^{\rm (Eq)}(T_{1};Q)}{2}\left.\frac{\partial}{\partial\Delta T}\Big[\mathcal{R}(T_{1},T_{1}+\Delta T;Q)\Big]\right|_{\Delta T=0}\Delta T\right],
\label{EnergyDeltaTTwoBaths}
\end{equation}
%
%
\begin{equation}
J_{k}(T_{1},T_{1}+\Delta T;Q)
\approx-4\Gamma_{k}\mathcal{W}E_{0}^{\rm (Eq)}(T_{1};Q)\mathcal{F}^{\rm (Eq)}(T_{1};Q)\left[1-\frac{\mathcal{F}^{\rm (Eq)}(T_{1};Q)}{2}\left.\frac{\partial}{\partial\Delta T}\Big[\mathcal{R}(T_{1},T_{1}+\Delta T;Q)\Big]\right|_{\Delta T=0}\Delta T\right].
\label{CurrentDeltaTTwoBaths}
\end{equation}

Now, to simplify these expressions, it would be useful to work out the equilibrium values and the derivative of the virial ratio. From Eq.(\ref{KVDoubleOccupation}), we can calculate the $E_{0}^{\rm (Eq)}(T_{1};Q)$ for the two baths scenario. Moreover, from the same equation we can calculate the energy for a single bath case $E_{0}^{(1)}(T_{1},Q_{1})$ where now the quality factor is given by $Q_{1}\equiv\Omega/(4\Gamma_{1})$. Notice that connection between the single quality factor $Q_{1}$ and the total one $Q$ is given by $Q=\gamma Q_{1}$. From the comparison of both expressions, immediately it is proven that:
\begin{equation}
E_{0}^{\rm (Eq)}(T_{1};Q)=E_{0}^{(1)}(T_{1},Q).
\end{equation}

This means that the value of the energy at equilibrium for the two baths scenario characterized by a quality factor $Q$ corresponds to the value of the energy at the same temperature for the single bath case given that the quality factor of this oscillator also corresponds to $Q$. Thus, by adjusting the value of the damping $\Gamma_{1}$ to get a quality factor $Q$ in the single bath scenario, we can obtain the equilibrium value for the two baths case when they are at equilibrium. The same happens for the virial factor $\mathcal{F}^{\rm (Eq)}(T_{1};Q)=\mathcal{F}^{(1)}(T_{1};Q)$.

A similar connection can be proven for the derivative of the virial ratio $\mathcal{R}$ for the two baths scenario. For the single bath case, employing Eq.(\ref{KVDoubleOccupation}) and considering that $\mathcal{R}=\langle K\rangle/\langle V\rangle$, we can prove first that for the single bath scenario:
\begin{equation}
\frac{\partial\mathcal{R}^{(1)}}{\partial T_{1}}(T_{1};Q_{1})=\frac{1}{T_{1}}\frac{\int_{0}^{+\infty}\frac{du}{\pi}\int_{0}^{+\infty}\frac{du'}{\pi}
\frac{u\left[u'^{2}f\left(\frac{u'}{u_{\rm C}}\right)-u^{2}f\left(\frac{u}{u_{\rm C}}\right)\right]u'}{\left([1-u^{2}]^{2}+\frac{u^{2}}{Q_{1}^{2}}\right)\left([1-u'^{2}]^{2}+\frac{u'^{2}}{Q_{1}^{2}}\right)}\coth\left(\frac{u}{\widetilde{T}_{1}}\right)\frac{u'}{\widetilde{T}_{1}}{\rm csch}^{2}\left(\frac{u'}{\widetilde{T}_{1}}\right)}{\int_{0}^{+\infty}\frac{du}{\pi}\int_{0}^{+\infty}\frac{du'}{\pi}
\frac{uu'}{\left([1-u^{2}]^{2}+\frac{u^{2}}{Q_{1}^{2}}\right)\left([1-u'^{2}]^{2}+\frac{u'^{2}}{Q_{1}^{2}}\right)}\coth\left(\frac{u}{\widetilde{T}_{1}}\right)\coth\left(\frac{u'}{\widetilde{T}_{1}}\right)},
\end{equation}
which for the two baths scenario, it allow us to prove that:
\begin{equation}
\left.\frac{\partial}{\partial\Delta T}\Big[\mathcal{R}(T_{1},T_{1}+\Delta T;Q)\Big]\right|_{\Delta T=0}=(1-\gamma)\frac{\partial\mathcal{R}^{(1)}}{\partial T_{1}}(T_{1};Q),
\end{equation}
connecting the derivative to the single bath scenario. From Fig.\ref{FigVirial}, we can observe that a maximum value of the derivative happens around $\widetilde{T}_{1}\sim 1-2$, giving a natural selection for the reference temperature of the ancilla bath.

All in all, we have connected all the quantities of the two baths scenario appearing in Eqs.(\ref{EnergyDeltaTTwoBaths}) and (\ref{CurrentDeltaTTwoBaths}) in terms of measurable quantities of the single case scenario. Therefore, from measurements on the single bath case, taking the bath as the ancilla bath, and one measurement either on the energy of the oscillator or the current flowing to the ancilla bath, it is possible to determine the temperature of the target bath without having access to it, and with measurements that can be amplified despite the temperature difference, assumed to satisfy $\Delta T\ll T_{1}$. For each case, the latter is given by:
\begin{equation}
\frac{E}{E_{0}^{\rm (1)}}\approx\mathcal{W}\mathcal{F}^{\rm (1)}\left[1-(1-\gamma)\frac{\partial\mathcal{R}^{(1)}}{\partial T_{1}}\frac{\mathcal{F}^{\rm (1)}}{2}\Delta T\right],
\label{DeltaTasE}
\end{equation}
\begin{equation}
\frac{J_{1}}{4\Gamma_{1}E_{0}^{\rm (1)}}\approx\mathcal{W}\mathcal{F}^{\rm (1)}\left[(1-\gamma)\frac{\partial\mathcal{R}^{(1)}}{\partial T_{1}}\frac{\mathcal{F}^{\rm (1)}}{2}\Delta T-1\right],
\label{DeltaTasJ1}
\end{equation}
where we have removed the dependences on each quantity just for simplicity at this point.

Given these results, a protocol for thermometry of the target bath gets clear. Either the protocol is based on energy or ancilla's current measurements, there are a few common assumptions, namely:
\begin{itemize}
\item \emph{Determining $\Omega$ and $\Gamma_{1}$:} A measurement of the decay of the oscillator when it is only coupled to the ancilla bath and no frequency noise is applied ($D=0$), is performed to obtain both its natural frequency $\Omega$ and the decay $\Gamma_{1}$.
\item \emph{Determining $\Gamma_{2}$:} Next, a measurement of the decay of the oscillator when it is coupled to both the ancilla and the target baths is performed to obtain $\Gamma_{2}$.
\item \emph{Single bath quantities:} The oscillator is just coupled to the ancilla bath. The damping of the ancilla bath is adjusted to have an oscillator with the total quality factor $Q$ (and not with the single bath quality factor $Q_{1}$).
\begin{itemize}
\item \emph{Energy measurement without frequency noise:} A energy measurement is assumed to be possible when frequency noise is not applied ($D=0$), obtaining $E_{0}^{(1)}$.
\item \emph{Measurements while applying frequency noise}: While applying frequency noise ($D\neq 0$), measurements are applied in order to determine $\mathcal{R}^{(1)}$ around values $\widetilde{T}_{1}\sim 1-2$, obtaining $\mathcal{F}^{(1)}$ and $\partial\mathcal{R}^{(1)}/\partial T_{1}$.
\end{itemize}
\item \emph{Two bath measurement:} The oscillator is coupled to both the ancilla and target baths with a total quality factor $Q$. An energy (current) measurement is performed while frequency noise is applied ($D\neq 0$) and the value is suitably magnified, obtaining $E$ ($J_{1}$).
\end{itemize}

Subsequently, $\Delta T$ is obtained from either Eq.(\ref{DeltaTasE}) or Eq.(\ref{DeltaTasJ1}) on repeated measurements for a chosen value of $D$, or alternatively from obtaining the slope when $D$ (and, then, $\mathcal{W}$) is varied. Every strategy presents the remarkable advantage that no direct access or intervention of the target bath is performed or required.

\end{document}